\documentclass[a4paper,oneside,11pt]{article}

\usepackage[utf8]{inputenc} 
\usepackage[T1]{fontenc} 
\usepackage{lmodern} 
\usepackage[a4paper]{geometry}
\usepackage{graphicx}
\usepackage{caption}
\usepackage{csquotes}
\usepackage{subcaption}
\usepackage{pdfpages}
\usepackage[flushleft]{threeparttable}
\usepackage{footnote}
\makesavenoteenv{array}
\usepackage[bottom]{footmisc}
\usepackage{tabularx, booktabs}
\newcolumntype{Y}{>{\centering\arraybackslash}X}
\newcolumntype{C}[1]{>{\centering\let\newline\\\arraybackslash\hspace{0pt}}m{#1}}
\usepackage{longtable}
\usepackage{titlesec}
\titleformat{\chapter}[hang]{\bf\huge}{\thechapter}{2pc}{}
\titlespacing*{\chapter}{0pt}{-1.5cm}{20pt}
\usepackage{sidecap}
\usepackage{pdflscape}
\usepackage{wrapfig}
\usepackage{listings}
\usepackage[toc,page]{appendix}
\usepackage{lipsum}
\usepackage{pgfplots}
\usepackage{amsmath} 
\usepackage{amssymb} 
\usepackage{esvect} 
\usepackage[amssymb]{SIunits}
\usepackage{eurosym}
\usepackage{mathrsfs}
\usepackage{dcolumn}
\usepackage{pifont}
\usepackage{amsthm}
\usepackage{hyperref}
\usepackage{datetime} 
\usepackage{bbm}

\newtheorem{theorem}{Theorem}

\pgfplotsset{compat=1.11}

\begin{document}

\begin{titlepage}
\title{Lorenz curves interpretations of the Bruss-Duerinckx theorem for resource dependent branching processes}
\author{Alexandre Jacquemain\thanks{Electronic address: \texttt{alexandre.jacquemain@uclouvain.be}}\\
  \multicolumn{1}{p{.7\textwidth}}{\centering\emph{Université catholique de Louvain, 1348, Louvain-la-Neuve, Belgium}}}
\date{\today}
\maketitle
\begin{abstract}
\noindent The Bruss and Duerinckx theorem for resource dependent branching processes states that the survival of any society form is nested in an envelope formed by two extreme policies. The objective of this paper is to give a novel interpretation of this theorem through the use of Lorenz curves. This representation helps us visualize how the parameters interplay. Besides, as we will show, it clarifies the impact of inequality in consumption. \\
\vspace{0in}

\bigskip
\end{abstract}
\setcounter{page}{0}
\thispagestyle{empty}
\end{titlepage}
\pagebreak \newpage

\section{Introduction}

Bruss and Duerinckx (2015) \cite{BrussDuerinckx2015} model the development of human populations and the influence of society forms through so-called resource dependent branching processes (RDBPs)\footnote{For a more general reference on branching processes, the authors refer to Haccou, Jagers and Vatutin (2007) \cite{Hacou2007}.}. Throughout this paper, we focus solely on RDBPs and borrow the notations of \cite{BrussDuerinckx2015}.

In short, RDBPs are special models of branching processes in which individuals create and consume resources. A so-called policy (society rule) determines how resources are distributed among the present individuals, and individuals have a means of interaction. RDBPs are models in discrete time with asexual reproduction. These assumptions are made for simplicity.
It should be pointed out, however, that the hypothesis  of asexual reproduction, which may first seem inadequate, is  justified by the notion of \textit{averaged reproduction rate of mating units} defined in  Bruss (1984) \cite{Bruss1984}. Indeed, Theorem 1 of the latter reference shows that for all relevant long-term questions concerning possible survival and equilibria, the simplification is asymptotically perfectly in order.

The remaining of this section is first dedicated to introduce formally RDBPs (section \ref{Section:RDBP}) and then to present the concept of Lorenz curves (section \ref{Section:LC}). In section \ref{Section:WFS_SFS}, we examine the contribution of Lorenz curves for analyzing two specific policies, the weakest-first and strongest-first societies. Section \ref{Section:Env} provides insightful interpretations of the envelopment theorem. Further motivation to introduce Lorenz curves is provided in section \ref{Section:immigration} which introduces immigration. Finally, section \ref{Section:ccl} concludes. 

\subsection{Resource dependent branching processes}
\label{Section:RDBP}

In the model of Bruss and Duerinckx, the two natural hypotheses and driving forces are
\begin{itemize}
\item Hyp.1: Individuals want to survive and see a future for their descendants, and
\item Hyp.2: In general, individuals prefer a higher standard of living to a lower one,
\end{itemize}
where Hyp.1. takes priority before Hyp.2, if these are incompatible. These hypotheses are modeled through a branching process revolving around four main aspects: reproduction, resources, claims and policies. 

\paragraph*{\textit{Reproduction}}
~~\\
Besides the simplification that reproduction is asexual, it is also assumed that all individuals reproduce independently of each other and follow the same reproduction law $(p_j)_j$, where $p_j$ denotes the probability that one individual has $j$ descendants. In order to avoid trivial cases, we suppose $p_0>0$ and $p_j$ for at least some $j>1$.

The reproduction matrix $(D_n^k)_{n,k}$ gathers i.i.d random variables $D_n^k$ representing the number of descendants of the kth individual at the nth generation. Given what we previously stated, $P(D_n^k=j)=p_j$. Besides, $m \equiv E[D_n^k]<\infty$ is the average number of offspring. Finally, $D(k)=D_n(k)\equiv\sum_{j=1}^k D_n^j$ is the total number of descendants of the nth generation given that this generation counts $k$ individuals. Note that the $n$ index can be omitted since we are dealing with i.i.d variables. 

\paragraph*{\textit{Resource space}}
~~\\
The resource space is viewed as a common pot that society distributes among its members and is made of heritage, to which we add individual production and subtract individual consumption. 

The resource creation matrix $(R_n^k)_{n,k}$ gathers all i.i.d individual resource creations $R_n^k$ and $r \equiv E[R_n^k] < \infty$ is the average productivity of an individual. Finally, $R(k)=R_n(k)\equiv\sum_{j=1}^k R_n^j$ is the total resource space.

\paragraph*{\textit{Claims}}
~~\\
In RDBPs, individual have a means of interactions through claims. The basic idea is that an individual stays in the society only if his claim is met, otherwise he leaves.

The claim matrix $(X_n^k)_{n,k}$ gathers all i.i.d individual claims $X_n^k$. Besides $F(x)=P(X_n^k\leq x)$ is the distribution of the claims and $\mu \equiv E[X_n^k]$ is the average claim. 

\paragraph*{\textit{Society}}
~~\\
A policy $\pi$ is defined as any function determining a priority order in the society. Individual claims are then met respecting this order until the resource space is exhausted. 

\paragraph*{\textit{RDBP}}
~~\\
A RDBP is any counting process $(\Gamma_t)_t$ with an initial ancestor $\Gamma_0=1$ and for which the population at the next period is determined by 
\begin{align*}
\Gamma_{n+1} & = Q^\pi(D(\Gamma_n),(X_1,\ldots,X_{D(\Gamma_n)}),R(\Gamma_n))
\end{align*}
where $Q^\pi$ is the counting process based on the policy $\pi$. For further details we refer to \cite{BrussDuerinckx2015}. For a more detailed motivation of \cite{BrussDuerinckx2015} and implications of socio-economic interest we refer to  Waijnberg (2014) \cite{Wajnberg2014} and Bruss (2016) \cite{Bruss2016}. 
~~\\

As we will elaborate later on, the following condition plays a central role in \cite{BrussDuerinckx2015}: the population cannot survive forever unless
\begin{align}
\label{Eq:(*)}
mF(\tau) \geq 1
\end{align}
where $F$ is the continuous cumulative distribution function (CDF) of the random resource claim (consumption) of an individual, and $\tau$ the unique solution of the implicit equation
\begin{align}
\label{Eq:(**)}
m \int_0^\tau x dF(x) = r
\end{align}
Equation (\ref{Eq:(*)}), which will attract our interest in the present work, showed up in a more general context already in Bruss and Robertson (1991) \cite{Bruss1991}, and, in a further extension, in Steele (2016) \cite{Steele2016} but we will have an innovative look at it. Our interest is to rephrase and re-interpret this condition in terms of the well-known notion of \textit{Lorenz curve}. 

\subsection{Lorenz curves}
\label{Section:LC}

Lorenz curves have been extensively used in the context of income distributions to portray how the proportion of total income owned by the up-to-$p$ poorest individuals evolve with $p$. Though it was introduced as soon as 1905 by the American economist Max Lorenz \cite{Lorenz1905} in order to picture social inequalities, it gained much more attention with the work of Atkinson (1970) \cite{Atkinson1970}, which provided a normative rationale for the use of Lorenz curves to measure inequality. 

For the definition of the Lorenz curve, we follow Gastwirth (1971) \cite{Gastwirth1971}. The Lorenz curve of $F$ at ordinate $p$ is defined as 
\begin{align*}
LC(p) & \equiv \frac{1}{\mu}\int^p_0 F^{-1}(t)dt
\end{align*}
Proceeding to the following change of variable $t=F(x)$, we can rewrite the Lorenz curve at $p$ as: 
\begin{align*}
LC(p) & = \frac{1}{\mu}\int_0^{F^{-1}(p)}xf(x)dx = \frac{1}{\mu}\int_0^{F^{-1}(p)}xdF(x)\\
& = \frac{1}{\mu}E\left[XI[F(X)\leq p]\right]
\end{align*}
In terms of properties, the LC passes through $(0,0)$ and $(1,1)$, is always increasing\footnote{This is true provided that the variable of interest is nonnegative. This is not true in full generality.} and convex.

\begin{figure}[!h] 
\centering 
\begin{tikzpicture}[scale=.8]
\begin{axis}[
	xlabel={Cumulative share of the population},
    ylabel={Cumulative share of the variable},
	xtick={0,1},ytick={0,1},
	xmin=-0.05,xmax=1.05,ymin=-0.05,ymax=1.05]    
    ]
    \draw[rounded corners=6ex] (0,0) -- (0.4,0.1) -- (0.8,0.4) -- (1,1);
    \draw[dashed] (0,0) -- (1,0);
    \draw[dashed] (1,0) -- (1,1);
    \draw[dotted] (0,0) -- (1,1);
    \node at (axis cs:  .48,  0.25) {$L(p)$};
\end{axis}
\end{tikzpicture}
\caption{Representation of a Lorenz curve (solid) as well as perfect equality (dotted) and perfect inequality (dashed)} 
\label{Fig:LCplot}
\end{figure}
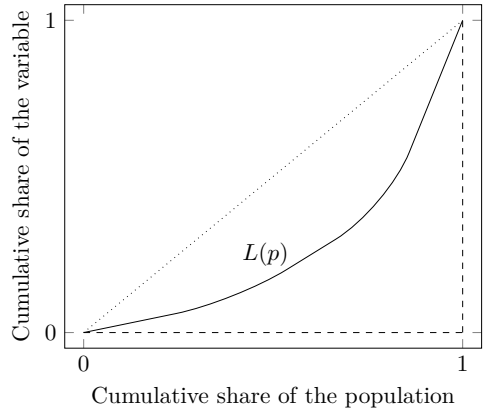

Turning to intuition, the Lorenz curve answers the following question: what share of claims do the up-to-$p$ most modest individuals gather? If the curve is a straight line, we call it the \textit{line of equality} (LOE): the bottom $5\%$ gather $5\%$ of the mass of claims, the $10\%$ most modest gather $10\%$, and so on. Hence we have perfect equality. If the curve is right-angle shaped, all the claims are spoiled by the most demanding individual, this is perfect inequality. Figure \ref{Fig:LCplot} displays a typical Lorenz curve (dotted) as well as the situations of perfect equality (solid) and perfect inequality (dashed).
~~\\

We now have the necessary tools to define \textit{Lorenz dominance}. We will say that $F_1$ \textit{LC-dominates} $F_2$ if the following holds
\begin{align*}
LC_1(p) \geq LC_2(p) \qquad \forall p \in [0,1]
\end{align*}
with strict inequality at some $p$. Atkinson (1970) \cite{Atkinson1970} provides normative reasons to use LC-dominance in order to rank societies in terms of inequality. The underlying idea is that all sensible inequality indices will deem $F_1$ to bear less inequality than $F_2$ provided that the Lorenz curve of $F_1$ is above that of $F_2$. If instead the two curves cross, one can find at least one pair of sensible inequality indices giving contradicting conclusions. The precise meaning of ``sensible'' is given in the latter reference. Obviously, the LOE LC-dominates all other Lorenz curves. 

The author is well aware that the interpretation of Lorenz curves in the context of RDBPs is in fact different from its usual sense.  The latter is that the LC was originally intended to measure the share of total wealth in a given state as a function of the share of its effectives. In condition \ref{Eq:(*)} for survival of a RDBP, it will intervene as the Lorenz curve of claims, which is not the same as present wealth, but rather as consumption. However, the link bewteen these notions is sufficiently close. See also Waijnberg (2014) \cite{Wajnberg2014} for a free interpretation.

\section{Weakest- and strongest-first societies}
\label{Section:WFS_SFS}

Bruss and Duerinckx (2015) \cite{BrussDuerinckx2015} present two extreme policies, the weakest-first and the strongest-first societies. These policies present a particular interest as they will form an envelope for the survival of any society in a sense we will make clear in section \ref{Section:Env}. 

\subsection{Weakest-first society (WFS)}
\label{Section:WFS}

The weakest-first society (WFS) serves the less demanding first. Formally, it can be defined through its counting process.
\begin{align*}
N(t,s) & = \begin{cases}
0 & \text{if } X_{<1,t>}>s\\
sup\{1\leq k\leq t: \sum_{j=1}^k X_{<j,t>} \leq s\} & \text{otherwise}
\end{cases} 
\end{align*}
where $X_{<j,t>}$ is the jth order statistics. The RDBP associated to the WFS is $(W_n)_n$ with $W_0=1$ and 
\begin{align*}
W_{n+1} & = N(D(W_n),(X_1,\ldots,X_{D(W_n)}),R(W_n))
\end{align*}

An important challenge related to RDBPs lies in determining under which conditions society can survive, given a particular policy. If the population is sure to asymptotically die out, Hyp.1. kicks in and society has to change the policy. More specifically, we want to determine when $q_\Gamma = P(\lim_{n\rightarrow \infty}\Gamma_n=0|\Gamma_0=1)$ is exactly equal to one. Theorem \ref{thm-Extinction_WFS} answers this question in the case of the WFS. 
\begin{theorem}[Extinction criterion for the WFS]
\label{thm-Extinction_WFS}
Let $(W_n)_n$ be the weakest first process on $(D_n^k,X_n^k,R_n^k)_{n,k}$
\begin{itemize}
\item[(a)] If $r\leq m\mu$ and $\tau$ is the solution of $\int_0^{\tau}xdF(x) = r/m$, then 
\begin{itemize}
\item[(i)] $mF(\tau) < 1 \Rightarrow q_W = 1$ 
\item[(ii)] $mF(\tau) > 1 \Rightarrow q_W < 1$
\end{itemize}
\item[(b)] If $r>m\mu$, then $q_W<1$
\end{itemize}
\end{theorem}
Some features can be pointed out.
\begin{itemize}
\item Starting from a sufficiently large number of ancestors, $q_W < 1$ is substantially the same as $q_W = 0$ (see proposition 4.2. in \cite{BrussDuerinckx2015}). 
\item Case $(b)$ corresponds to situations where there are more resources than claims on average. As such, it is not interesting from a macroeconomical point of view.
\item $F(\tau) = 1/m$ is the threshold between almost sure extinction and possible survival of the WFS. 
\end{itemize}

We can use Lorenz curves to rewrite the conditions of this theorem. Note that throughout the following sections, we focus on the (interesting) case where $r \leq m\mu$. We have
\begin{align*}
\int_0^{\tau}xdF(x) = \frac{r}{m} & \Rightarrow \int_0^{F(\tau)}F^{-1}(t)dt = \frac{r}{m} \Rightarrow LC[F(\tau)] = \frac{r}{m\mu}\\
& \Rightarrow F(\tau) = LC^{-1}\left(\frac{r}{m\mu}\right)
\end{align*}
In order to have extinction of the WFS, one needs
\begin{align*}
mF(\tau) < 1 & \Leftrightarrow LC\left(\frac{1}{m}\right) > \frac{r}{m\mu}
\end{align*}
This is already an interesting result in itself. In order to settle about the extinction of the WFS, one needs not know $F$. It is sufficient to know $r,m,\mu$ as well as inequality, through the Lorenz curve. 

\subsection{Strongest-first society (SFS)}
\label{Section:SFS}

The strongest-first society (SFS) follows the opposite logic, it serves the most demanding first. Its counting process is defined as
\begin{align*}
M(t,s) & = \begin{cases}
0 & \text{if } X_{<t,t>}>s\\
sup\{1\leq k\leq t: \sum_{j=t-k+1}^t X_{<j,t>} \leq s\} & \text{otherwise}
\end{cases} 
\end{align*}
and its associated RDBP is $(S_n)$, with $S_0=1$ and 
\begin{align*}
S_{n+1} & = M(D(S_n),(X_1,\ldots,X_{D(S_n)}),R(S_n))
\end{align*}
In a way similar to before, theorem \ref{thm-Extinction_SFS} examines the survival pattern of the SFS. 
\begin{theorem}[Extinction criterion for the SFS]
\label{thm-Extinction_SFS}
Let $(S_n)_n$ be the strongest first process on $(D_n^k,X_n^k,R_n^k)_{n,k}$
\begin{itemize}
\item[(a)] If $r\leq m\mu$ and $\theta$ is the solution of $\int_{\theta}^{\infty}xdF(x) = r/m$, then 
\begin{itemize}
\item[(i)] $m[1-F(\theta)] < 1 \Rightarrow q_S = 1$ 
\item[(ii)] $m[1-F(\theta)] > 1 \Rightarrow q_S < 1$
\end{itemize}
\item[(b)] If $r>m\mu$, then $q_S<1$
\end{itemize}
\end{theorem}

The same comments as before can be made. Let's highlight the last one: $F(\theta) = 1 - 1/m$ corresponds to the threshold between extinction and survival in the SFS. Once again, we can use the Lorenz curve to rewrite the conditions of this theorem. We have
\begin{align*}
\int_{\theta}^{\infty}xdF(x) = \frac{r}{m} & \Rightarrow \mu - \int_0^{F(\theta)}F^{-1}(t)dt = \frac{r}{m}\\
& \Rightarrow 1-LC[F(\theta)] = \frac{r}{m\mu}\\
& \Rightarrow F(\theta) = LC^{-1}\left(1-\frac{r}{m\mu}\right)
\end{align*}
In order to have survival of the SFS, one needs
\begin{align*}
m[1-F(\theta)] > 1 & \Leftrightarrow LC\left(1-\frac{1}{m}\right) > 1-\frac{r}{m\mu}
\end{align*}
Similarly to the WFS, one needs not know the precise form of $F$ to determine whether the SFS can survive.

\section{Envelopment theorems}
\label{Section:Env}

As the WFS and SFS form extreme policies, one could wonder about the general relevance of results concerning them. Actually, it appears that they form an envelope for the survival of any society. This idea is formally expressed in Theorem \ref{thm-Envelopment}.

\begin{theorem}[Unconditional envelopment theorem]
\label{thm-Envelopment}
Assume $m[1-F(\theta)] \neq 1$ if $r \leq m\mu$. Then, the following holds
\begin{itemize}
\item Let $\Gamma_n(L)$ be any RDBP, $W_n(L)$ the WF-process and $S_n(L)$ the SF-process, all started with $L>0$ individuals, then 
\begin{align*}
P\left(\lim_{n\rightarrow \infty}S_n(L) \leq \lim_{n\rightarrow \infty}\Gamma_n(L) \leq \lim_{n\rightarrow \infty}W_n(L)\right) & \xrightarrow{L\rightarrow \infty} 1
\end{align*}
\item $q_W = 1 \Rightarrow q_{\Gamma} = 1 \Rightarrow q_S = 1$, for all RDBP $\Gamma$.
\item $q_S < 1 \Rightarrow q_{\Gamma} < 1 \Rightarrow q_W < 1$, for all RDBP $\Gamma$.
\end{itemize}
\end{theorem}

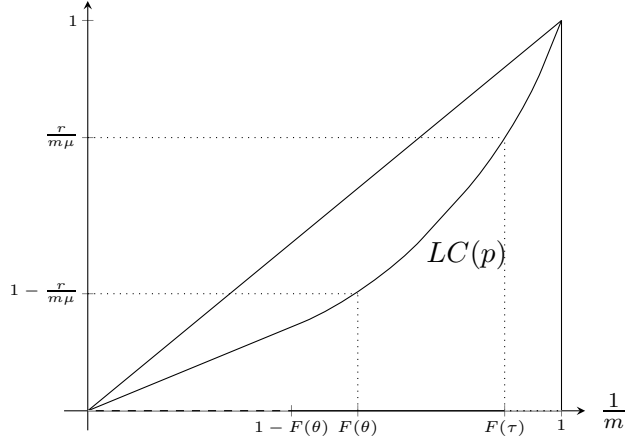
\begin{figure}[!h] 
\centering 
\begin{tikzpicture}[scale=1]
\begin{axis}[
	axis lines=middle,
	xlabel=$\frac{1}{m}$, xlabel style={at=(current axis.right of origin),xshift=0.4ex, anchor=west},
    ylabel=, ylabel style={at=(current axis.above origin),yshift=11ex, anchor=south},
	xtick={0,0.43,0.57,0.88,1},ytick={0,0.3,0.7,1},
	xticklabels={0,$1-F(\theta)$,$F(\theta)$,$F(\tau)$,1},xticklabel style = {font=\tiny,yshift=0.5ex},
	yticklabels={0,$1-\frac{r}{m\mu}$,$\frac{r}{m\mu}$,1},yticklabel style = {font=\tiny,xshift=0.5ex},
	xmin=-0.05,xmax=1.05,ymin=-0.05,ymax=1.05]    
    ]
    \draw[rounded corners=6ex] (0,0) -- (0.6,0.3) -- (0.9,0.7) -- (1,1);
	\draw[dotted] (0,.7) -- (.88,.7);
	\draw[dotted] (0.88,0) -- (0.88,.7);    
    \draw[dotted] (0,.3) -- (0.57,.3);
    \draw[dotted] (0.57,0) -- (0.57,.3);
    \draw (0,0) -- (1,0);
    \draw (1,0) -- (1,1);
    \draw (0,0) -- (1,1);
    \draw[dashed,white] (0,0)--(0.43,0);
    \draw[dash pattern=on 2pt off 0.5pt,white] (0.88,0)--(1,0);
    \node at (axis cs:  .8,  0.4) {$LC(p)$};
\end{axis}
\end{tikzpicture}
\caption{On the connection between the Lorenz curve and the Envelopment theorem} 
\label{Fig:LC_Envelopment}
\end{figure}

We can interpret the last two results as follows. $\Gamma_n$ will die out almost surely if the WFS does so and $\Gamma_n$ can survive if the SFS can survive. From the past section, recall that
\begin{align*}
F(\tau) & = LC^{-1}\left(\frac{r}{m\mu}\right) \qquad F(\theta) = LC^{-1}\left(1-\frac{r}{m\mu}\right)
\end{align*}
Besides, the WFS extincts if $F(\tau)<1/m$ and the SFS survives if $1-F(\theta)>1/m$. Figure \ref{Fig:LC_Envelopment} represents the context of the envelopment theorem. Note that $1/m \in [0,1]$ and, hence, can be represented on the horizontal axis.  
Focusing our attention to the horizontal axis, two regions are of particular interest. 
\begin{itemize}
\item Dashed area: it's the set of all $m$ such that $1/m < 1-F(\theta)$. Consequently, it is the set of all $m$ such that the society survives for sure (since the SFS survives as well). 
\item Dotted area: it's the set of all $m$ such that $1/m > F(\tau)$. Consequently, it is the set of all $m$ such that the society extincts for sure (since the WFS extincts as well). 
\end{itemize}
In other words, these two areas correspond to the set of all $m$ for which we know for sure that the society will finally survive or die out. Outside these areas, the answer will depend on the chosen policy. Now that this is understood, we can look at the effects of changing the parameters. 
\begin{enumerate}
\item \underline{What happens if $r$ increases or $\mu$ decreases}. Looking at the graph, we see that the dotted area decreases while the dashed area increases. This makes sense: more resources and less claims lead to a higher chance of survival.
\item \underline{What happens if inequality decreases}. Here we technically mean that the new Lorenz curve is everywhere above the other. In this case, it is easy to see that $F(\tau)$ and $F(\theta)$ both decrease. As such, both the dotted and dashed areas increase. This is interesting. Inequality doesn't have a clear-cut impact on the chance of survival of the society. Less inequality actually enlarges both the set of $m$ for which we have survival for sure and the set for which we have extinction for sure\footnote{Note in passing that it also means that less inequality increases the chance of survival in the SFS and decreases it in the WFS.}. What is striking is that less inequality undoubtedly increases the set of $m$ for which we are sure about the fate of society. In some sense, the less inequality, the less likely we need to know the policy to determine whether society will finally survive or die out.
\item \underline{Perfect equality}. Recall that the Lorenz curve in presence of perfect equality is the $45^{\circ}$ line. In such scenario, $F(\tau) = 1-F(\theta) = r/(m\mu)$. Hence, the dotted and dashed areas cover the entirety of the $[0,1]$ segment. Consequently, whatever the policy, we know society will finally die out or survive. 
\item \underline{Perfect inequality}. Recall that the Lorenz curve in presence of perfect inequality is triangle shaped. In this case, $F(\tau) = 1$ and $F(\theta) = 0$ and both areas vanish. Consequently, without knowing the underlying policy, there is nothing we can say about whether society will finally survive or die out. 
\end{enumerate}

\section{Introduction of immigration}
\label{Section:immigration}

In a recent special course on RDBPs taught by Bruss at the Université catholique de Louvain it was seen that Lorenz curves also intervene whenever one brings immigration into the picture and want to understand under which conditions the immigrant-population and the home-population can reach an equilibrium\footnote{The equilibrium is meant as the asymptotic ratio $\alpha$ between the size of the immigrant-population compared to the home-population.}, both conditioned on survival. We confine here our interest to the simplest case where from some finite time onward there are no new immigrants, and where the immigrants do not integrate into the home-populatoin. In this case, the necessary condition for  an equilibrium to exist can be expressed in terms of the condition
\begin{align}
\label{Eq:(***)}
m_h F_h(\tau) & = m_i F_i(\tau) \geq 1
\end{align}
where $\tau$ is solution of 
\begin{align}
\label{Eq:(****)}
m_h \int_0^\tau x dF_h(x) + \alpha m_i \int_0^\tau x dF_i(x) & = r_h + \alpha r_i
\end{align}
and where $F_h$ (respectively $F_i$) denote the CDF of claims for individuals from the home-population (respectively immigrant-population). Besides, $r_h$ ($r_i$) and $m_i$ ($m_h$) are the corresponding parameters of the two subpopulations. Note that the value $\tau$ is  the same in both integrals. Although we do not discuss equilibria of  home-population and immigrant-population in the present short article, we wanted to mention  the conditions (\ref{Eq:(***)}) and (\ref{Eq:(****)}) for a Bruss-equilibrium already in this present article. Indeed, this adds very much to our motivation to look at Lorenz curves under several angles of view,  because (\ref{Eq:(***)}) and (\ref{Eq:(****)}) must be satisfied simultaneously, and the condition $m_1F_1(\tau) = m_2F_2(\tau)$ is remarkably demanding. How we could interpret these conditions in a single combined Lorenz curve graphic remains a challenge. 

\section{Conclusion}
\label{Section:ccl}

In the context of RDBPs, the Bruss and Duerinckx theorem states that the chance of survival of any society is finally nested between two extreme cases, the weakest-first and the strongest-first societies. As already pointed out in Bruss and Duerinckx (2015) \cite{BrussDuerinckx2015}, this envelope is impacted by the individual productivity, the multiplication rate of the population and the distribution of claims. While the two first are simple parameters, the impact of the latter is more difficult to grasp. 

The contribution of the Lorenz curve is to disentangle the effect of the mean claim from its pure distributional aspects (i.e. inequality). By doing so, we observe that inequality in consumption has a clear-cut impact on the stringency of the envelope. Less inequality doesn't necessarily translate into more or less chance of survival, it yields more certainty. In practice, it means that less inequality increases the number of situations for which we are sure whether the society finally survives or dies out. We find this to be a remarkable property. 

Finally, note that the role of Hyp 2. listed in the introduction is only mentioned indirectly here.  Hyp 1. has priority, but within this limitation the population is then likely to increase the standard of living so that the long-term multiplication factor $mF(\tau)$ is likely to be chosen close to one, i.e. close to so-called criticality. RDBPs cannot be directly compared with Galton-Watson processes (as in Bruss (1978) \cite{Bruss1978}), or BPs in a random environment, but there are some interesting links of results around criticality. In his  course at UCL (2017), Bruss referred for further details to Jagers and Klebaner (2004) \cite{Jagers2004} and Afanasyev et al. (2005) \cite{Afanasyev2005}.

\end{document}